\documentclass{article}
\usepackage{array}
\usepackage{amsmath}
\usepackage{amssymb}        
\newcommand{\tick}{\checkmark}
\usepackage{amsfonts}       
\usepackage{arxiv}
\usepackage{booktabs}       
\usepackage{multirow}
\usepackage{multicol}
\usepackage{enumitem}
\setlist[enumerate]{leftmargin=0.7cm, label*=\arabic*.,itemsep=0pt}
\setlist[itemize]{itemsep=0pt} 
\newcounter{saveenum}
\usepackage{graphicx}
\usepackage{subcaption}
\graphicspath{ {./images/} }
\usepackage{hyperref}       
\usepackage{lipsum}
\usepackage{url}            
\usepackage{longtable}

\title{Exploring the Ethical Concerns in User Reviews of Mental Health Apps using Topic Modeling and Sentiment Analysis}

\author{
Mohammad Masudur Rahman \\
School of Computing and Informatics\\
University of Louisiana at Lafayette\\
Lafayette, LA, United States\\
\texttt{mohammad.rahman3@louisiana.edu} \\
\And
Beenish Moalla Chaudhry \\
School of Computing and Informatics\\
University of Louisiana at Lafayette\\
Lafayette, LA, United States\\
\texttt{beenish.chaudhry@louisiana.edu} \\
}

\begin{document}
\maketitle

\begin{abstract}
The rapid growth of AI-driven mental health mobile apps has raised concerns about their ethical considerations and user trust. This study proposed a natural language processing (NLP)-based framework to evaluate ethical aspects from user-generated reviews from the Google Play Store and Apple App Store. After gathering and cleaning the data, topic modeling was applied to identify latent themes in the context of ethics using topic words and then map them to well-recognized existing ethical principles described in different ethical frameworks; in addition to that, a bottom-up approach is applied to find any new and emergent ethics from the reviews using a transformer-based zero-shot classification model. Sentiment analysis was then used to capture how users feel about each ethical aspect. The obtained results reveal that well-known ethical considerations are not enough for the modern AI-based technologies and are missing emerging ethical challenges, showing how these apps either uphold or overlook key moral values. This work contributes to developing an ongoing evaluation system that can enhance the fairness, transparency, and trustworthiness of AI-powered mental health chatbots.
\end{abstract}
 

\section{Introduction}

Mental health disorders represent one of the most urgent and complex public health challenges in the word, affecting people across all regions, incomes, and cultures. In the United States alone, about one in five adults experiences a mental illness each year—amounting to over 60 million people in 2024 \cite{nih_data}. In the global context, the World Health Organization (WHO) reports that approximately 1 in 8 people, approximately 970 million individuals, live with some mental health condition \cite{who_data}. These statistics are severe enough to evident the impairment daily functioning of a large population, yet nearly 56\% of the people do not receive treatment due to the some barriers such as cost of medication, social stigma, and geographic disparities \cite{nih_data}. In the U.S., about 35\% of adults with mental health conditions report unmet needs for care, driven by financial hardship, fragmented healthcare systems, and cultural barriers \cite{apa_data}. While traditional therapy by medical professional has proven effective, it is still out of reach for many due to high costs and requires continuous visits to the medical which resposible for cost, time and consistency \cite{survey}. On top of that, societal stigma continues to deter people from seeking treatment making the issue more complex. In a 2023 survey, more than 60\% of respondents reported that shame or fear of judgment kept them away from accessing the professional care or treatment \cite{apa_data}.

In response to these barriers of traditional mental health care, AI-powered mental health mobile applications are rapidly emerging as an alternative that offers scalable, cost-effective and stigma-free solutions and make resources more accessible than ever. The rising global market for these mobile apps, valued at 1.27 billion dollars in 2023, is projected to increase exponentially to 8.47 billion dollars by 2032, glancing at a rate 23.5\% annually \cite{pharmiweb}. These apps leverage emerging technologies including artificial intelligence (AI), machine learning (ML), and natural language processing (NLP) to deliver personalized interventions such as mood tracking, cognitive-behavioral therapy (CBT) exercises, and human-like chat etc. For instance, mobile apps like Wysa or Woebot use NLP to create conversations based on CBT to help users manage sleep, stress, anxiety, and depression, while apps like Calm provide help with sleep, guided meditation, and relaxation exercises, managing stress and moods \cite{becky2028empathy, huberty2019efficacy}. These apps take the driving role as transformative tools for their 24/7 availability, accessibility to both rural and urban populations with limited access to clinicians, and low-income communities burdened by inequities. Moreover, these mental health apps are more affordable as many are free, while others cost only about 10-30 dollars per month which contrasts with conventional therapy democratizing access to mental health support. This lower cost not only make mental health support more accessible to a much wider population but also shows potential for reducing the financial burden on individuals seeking treatment. It makes the emotional support more inclusive and readily available. 

However, the lack of proper use of these mobile applications has sparked some serious ethical dilemmas in this context, such as can we even trust these tools to stick around for the long haul? These have threatened their long-term viability and trust issues. Because mental health data are sensitive, privacy and security have become significant concerns for users. Many apps gather sensitive information such as biometrics, behavior patterns, etc., often without user consent. Many of these apps do not fully comply with well-known data protection laws like HIPAA in the U.S. or GDPR in the EU, raising serious questions about how safely user information is being used. A 2023 investigation revealed that 72\% of mental health apps shared user data with third-party advertisers, often without explicit consent \cite{coghman23}. There rises some emegent concerns like algorithmic bias due to use of AI or predictive analytics that further adds another layer of risk, as many AI/ML models are trained on specific samples failing to represent diverse populations. This lack of inclusivity may lead to an unequal or even harmful outcomes to the non-represntative user groups reinforcing biases in the model. One study found that AI chatbots were 40\% less likely to recommend crisis support resources to users from minority ethnic backgrounds, revealing how biases in training data \cite{apa_data2}.

Moreover, transparency and accountability are still major challenges for these apps. Many apps are designed and implemented as a `black box' where users never know how the decision-making processes are made in those apps. These concerns are elevated as some apps are being available in the market without any clinical validation, which may cause critical issues when patients are blinded by the apps and trust the results to be real therapists. Some claim adherence to CBT principles, while fewer than 15\% have undergone rigorous randomized controlled trials (RCTs) to verify outcomes, a study revealed \cite{apa_data2}. On raising this type of evidence, mental health care professionals express concerns and skepticism. The expert organizations, like the American Psychological Association (APA), caution against overreliance on unvalidated tools for severe conditions like schizophrenia or suicidal ideation.

Despite these raising concerns in this age of AI, we can not avoid the usefulness incoporating AI in intervention f mental health solving problems rather we harness the potential of AI while mitigating harm. To serve the purpose, mental health stakeholders including regularity bodies, medical professionals, mobile app developers, and most importantly the users must adopt a multidisciplinary approach. Mobile app developers should give more emphasis on ethical principles such as fairness, accountability, user privacy etc. into their application design and implementation. These findings emphasize the need for a systematic framework to evaluate the ethical aspects of mental health apps. Such a framework would help the stakeholders (developers, researchers, and users) to understand whether these apps respect with principles of privacy, fairness, transparency, and user well-being, etc. It would also guide the ethical and responsible design of AI-based mental health tools, ensuring they genuinely help users in a safe, fair, and trustworthy way.

About 10,000 mobile apps for mental health are available in the Google Play Store and Apple App Store, a few of them are recognised as Software as a Medical Device (SaMD) by the US Food and Drug Administration (FDA) \cite{torous2025evolving}. Most of them are not rigorously assessed for ethical compliance \cite{babu2025digital}. Camacho et al. reviewed 578 mental health apps and exposed dicrepencies between app feastures and privacy valunabilities at simgnificant risks \cite{camacho2022assess}. Another study analysed 92 mobile apps on depression, anxiety and mood using the App Evaluation Model on five criteria: accessibility, integrity, clinical evidence base, user engagement and interoperability which revealed that only one of the them met all five criteria and only 3\% demonstrated sufficient privacy safeguards \cite{Nikki}. This App Evaluation Model by American Psychiatric Association provides a general framework for assessing mental health apps; only partially addresses ethical concerns rather than a comprehensive evaluation of ethical considerations \cite{app_model}.  So, it is important to ethically evaluate these apps to balance their growing demand with the responsibility of ensuring safety, privacy and effectiveness \cite{olawade2024enhancing}. Proper evaluations not only protect users but also help developers improve their apps by identifying possible ethical issues early in the design process.

\subsection{Methods for Assessing Ethical Concerns}

Evaluating the ethical considerations of digital mental health apps is critical to ensure they are safe, fair, and respectful to the end-users. Traditionally, this has been done through human expert reviews. In this process, mental health professionals set up some theoretical criteria which help to check whether an app protects privacy, ensures safety, treats users fairly, and takes responsibility for its actions \cite{mccormack2024}. There are some well-known guideline frameworks, such as the Belmont Report \cite{belmont}, the Declaration of Helsinki \cite{helsinki}, GDPR in the EU \cite{gdpr}, and HIPAA \cite{hipaa}. These evaluation frameworks are based on moral ideas like doing what leads to good outcomes and following ethical rules. However, since experts are human, their judgments can sometimes be biased or inconsistent, and they ignore the empirical perspective of end-users, which is important in this case. Moreover, these methods are difficult to scale for large numbers of apps and often require collaboration across multiple disciplines to provide a truly comprehensive assessment. These manual, rule-based approaches fail to capture what real users are saying. Users are among the primary and most significant stakeholders in mobile mental health applications, and their open, honest feedback can reveal important insights about their experiences and concerns. Thousands of users share their experience and feedback every day. These unguided, free-text review feedback give valuable insights into real user experiences. However, most existing ethical frameworks fail to consider what users say when evaluating app design and development. Manual assessments are also prone to human mistakes, especially when reviewing large amounts of data. As a result, they often miss subtle issues like fairness, bias leading to incomplete or inaccurate evaluations. So, these manual methods are less effective for understanding ethical concerns from the huge and constantly growing amount of user reviews.

\subsection{Purpose of this Paper}

This study evaluates how effectively mental health mobile applications adhere to ethical principles derived from user feedback. Following currently available ethical guidelines, it will focus on how users feel about their data privacy, informed consent, accessibility, and the transparency of the app interactions, specifically based on the use of emerging artificial intelligence in mobile apps. We emphasized user-generated open-text reviews rather than strict ethical concerns analysis guidelines. We focus on understanding these concerns from the user perspective and whether they can provide valuable insights into the ethical implications of these widely used digital tools for mental health care. The work followed a systematic approach, beginning with creating a dataset of publicly available user reviews from Google Play Store and Apple App Store. The dataset consisted of thousands of individual user reviews from five Cognitive behavioral therapy (CBT)-based mental health chatbots: Wysa, Woebot, Youper, Sintelly, and Elomia. Then, the reviews were analyzed using natural language processing techniques and aligned with existing ethical concerns and emergent ethics. The sentiment analysis part of the study reveals whether the reviews support or reject aligned ethical aspects. 

\subsection{Contributions of the Paper}

This paper presents a data-driven approach to exploring ethical issues in AI-powered mental health apps by analyzing open-text user reviews. It introduces a novel NLP-based framework for assessing ethical considerations and develops a large dataset to support this evaluation. Together, these efforts provide an evidence-based foundation for strengthening AI ethics in the rapidly evolving field of digital mental health care. The key contributions are outlined below:
\begin{enumerate}
    \item We build a large dataset of over 66,000 user reviews collected from different mental health apps. By organizing this data, we create a benchmark that helps measure how users perceive the ethical use of AI in mental health apps.
    \item The paper introduces a novel pipeline that utilizes natural language processing techniques to systematically assess ethical considerations. By leveraging topic modeling and sentiment analysis, the pipeline can identify concerns about privacy, data security, bias,  algorithm transparency, etc.
    \item We clearly define what ethical concerns mean in the context of AI-powered mental health apps, outlining their scopes and practical use cases based on recognized guidelines from governments and other authorities. By analyzing open-ended user reviews, we also identify important ethical issues that existing frameworks often overlook in current mobile mental health applications.
\end{enumerate}

\section{Existing Ethical Frameworks}

\subsection{Introduction to Ethical Frameworks}
Ethical considerations refer to the moral principles and standards that help us to decide what is right or wrong when doing research, application development, technology design, and human interaction. They serve as a framework to ensure respect for human dignity, minimize harm, promote fairness and transparency, and address other concerns. Ethical considerations in any form of research or innovation, especially those involving human data or decision-making systems, help maintain trust between developers and users by ensuring that societal and individual well-being remain central to the process \cite{deshpande2020ethical}. With the advent of artificial intelligence, it raises a need to guide responsible development and deployment of AI-based applications along with ethical regulations in the sensitive mental health domain. Ethical frameworks provide a structured approach to assessing ethical issues, ensuring that AI is designed and used to align human values and well-being, reduce risks, and build trust in AI systems. Such frameworks are often referred to as a term ``Responsible AI (RAI)" that involves incorporating ethical principles throughout the entire application development lifecycle, from design and testing to deployment and ongoing use, ensuring to reduce bias and stay focused on human-centric outcomes while balancing innovation with care for users \cite{diaz2023connecting, resAI}. Different countries and organizations have developed their own versions of ethical consideration framework linking the ethical principles to system requirements. Some of the ethical aspects are discussed here.

Human-centered values and empathy emphasize user dignity, compassion, and inclusivity in design, ensuring emotional safety and psychological support. Transparency and explainability require AI systems to provide a clear flow of how data is collected and decisions are processed to build user trust. Fairness, equity, and inclusivity prevent bias and ensure equal access across gender, race, culture, and socioeconomic status. Privacy, consent, and data protection safeguard personal information through clear consent and minimal data collection, reinforcing autonomy and control. Accountability and responsibility ensure traceability of outcomes and clarify who is answerable for unintended consequences. Non-maleficence and beneficence define the balance between avoiding harm and promoting user well-being, especially in sensitive domains like mental health.

We need a framework to evaluate ethical considerations because research, technology, or artificial intelligence ethics is complex, multidimensional, and context-dependent. A framework provides a structured, systematic way to identify, analyze, and manage ethical issues rather than relying on human intuition alone. Such frameworks bridge the gap between ethical theory and real-world practice, enabling responsible innovation that truly benefits society \cite{mccormack2024}.

\subsection{Existing Ethical Frameworks in AI and Healthcare}


Ethical consideration is defined as a set of principles and techniques that employ the standards of right and wrong to guide moral conduct in developing and designing AI technologies. Certain ethical principles related to healthcare and medical research can be employed in developing AI systems \cite{leslie2019}. These frameworks may differ in their goals and focus, but they all aim the same purpose: to protect human values, prevent harm, promote fairness, and ensure accountability in how technology is used. One of the earliest ethical guidelines, the Belmont Report (1979), introduced four key ethical principles: respect for autonomy, beneficence, nonmaleficence, and justice \cite{belmont}. The Belmont Report laid the foundation for many modern ethical frameworks used today, not only in medical research but also in technology and AI ethics. Its main principles are summarized below:
\begin{enumerate}
    \item Respect for Autonomy covers the principle of self-governance. Physicians and researchers respect the right of an individual to make their own decisions regarding healthcare and participation in research. In the case of AI, application developers should ensure that users have full control over their interactions with applications.
    \item Beneficence involves the principle of doing good. Healthcare professionals and researchers act for the benefit of the individual. Just like AI developers are responsible for user good.
    \item Nonmaleficence indicates the principle of avoiding harm. Physicians and researchers do not intentionally, negligently, or unintentionally harm the individual. In the context of AI, application designers are responsible for preventing harm and mitigating risks.
    \item Justice refers to the principle of fairness. Physicians and researchers should treat people fairly and equitably, regardless of race, gender, religion, socioeconomic status, or medical condition. Similarly, AI developers and users are responsible for ensuring equity and fairness, regardless of factors.
\end{enumerate}

Similar to the Belmont Report, the Declaration of Helsinki (DoH) was also another of the earliest foundations of ethical guidelines, established by the World Medical Association (WMA) for medical research involving human subjects \cite{doh}. It was first adopted in 1964 and subsequently revised most recently in 2024. Its core principle is protecting the health, well-being, and rights of human research participants, including principles on informed consent, risk-benefit assessment, equitable distribution of benefits, and community engagement, with recent updates addressing AI. The key principles are: 
\begin{enumerate}
    \item Well-being of human participant: The health and well-being of research participants must take precedence over the interests of science.
    \item Informed Consent: Participants must provide informed consent to participate in research.
    Risk and Benefit Assessment: Research participants should not be exposed to greater risk.
    \item Equitable Distribution: The benefits, risks, and burdens of research should be distributed fairly among participants and communities.
    \item Community Engagement: Researchers should engage with communities to understand their needs and goals and share research results with them.
    \item Scientific Integrity: The declaration includes principles to ensure the scientific rigor and integrity of research, with a zero-tolerance stance on misconduct.
\end{enumerate}

In 2017, during the Asilomar Conference on Beneficial AI, the Asilomar AI Principles were established to guide the development and deployment of AI technologies \cite{asilomar}. These principles offer a framework for ensuring that AI evolves and is safe, ethical, and beneficial to humanity. It consists of 23 guidelines grouped into three categories: Research Issues, Ethics and Values, and Longer-Term Issues. Here, we only focus on ethical aspects, which cover:
\begin{enumerate}
    \item Safety: AI systems must be safe and secure throughout their operational lifetime.
    \item Failure Transparency: If an AI system causes harm, it should be possible to determine why.
    \item Judicial Transparency: Autonomous systems involved in judicial decisions should provide explanations that humans can understand.
    \item Responsibility: Designers and developers of AI have a moral responsibility to consider its implications.
    \item Value Alignment: AI systems should align their goals and behaviors with human values.
    \item Human Control: Humans should decide how and when to delegate decisions to AI systems.
    \item Personal Privacy: Individuals should control the data they generate.
\end{enumerate}

IEEE released an ethical framework: Ethically Aligned Design (EAD) with recommendations and principles that guide the ethical development of autonomous and intelligent systems in 2016 \cite{ieee_ead}. It incorporates transparency, accountability, and human-centered design in autonomous and intelligent systems. The following ethical aspects are covered by IEEE EAD for healthcare applications:
\begin{enumerate}
    \item Human Rights: Ensuring systems respect any or all recognized human rights.
    \item Accountability: Signs, audits, and safeguards to hold designers and operators responsible for system impacts.
    \item Transparency: Systems should be explainable and their operations comprehensible to users.
    \item Well-Being: Aligning AI systems with metrics prioritizing human and social well-being.
\end{enumerate}

The OECD AI Principles, adopted in 2019, serve as the first intergovernmental standard for the development and governance of AI \cite{oecd}. These principles ensure that AI technologies are innovative, trustworthy, and aligned with human rights and democratic values. The framework is organized into two main sections: value-based principles and recommendations for policymakers. The OECD AI Principles complement five values to guide the ethical and responsible development and deployment of AI systems. These include the following:
\begin{enumerate}
    \item Inclusive Growth, Sustainable Development, and Well-Being: AI systems should enhance human capabilities, reduce inequalities, and promote environmental sustainability. Stakeholders are encouraged to responsibly steward AI technologies to augment creativity and include underrepresented populations
    \item Respect for human rights and democratic values: AI actors must uphold values like non-discrimination, fairness, and privacy throughout the AI lifecycle. Mechanisms to address misinformation amplified by AI, while respecting freedom of expression, are integral to this principle. 
    \item Transparency and Explainability: AI systems should provide meaningful information about their logic, behavior, and decision-making processes. Transparency enables users to understand and challenge outputs, fostering public trust.
    \item Robustness, Security, and Safety: The principle emphasizes that AI systems must function reliably in adverse conditions and safeguard users from unreasonable risks. Mechanisms to override, repair, or decommission AI systems posing undue harm are also promoted.
    \item Accountability: AI actors are obligated to ensure the proper functioning of their systems in alignment with ethical and technical standards. Key aspects include system traceability, risk management, and cooperative responsibility among stakeholders.
\end{enumerate}

In 1019, the High-Level Expert Group on Artificial Intelligence (HLEG) set up by the European Commission (EU) developed an Ethics Guideline for the Trustworthy AI framework consisting of three main components: lawfulness, ethics, and robustness \cite{eu_tai}. They designed the framework in three core steps: foundations, realizations, and assessment of trustworthy AI. The foundations identify and describe four ethical principles based on fundamental human rights: respect for human autonomy, harm prevention, fairness, and explicability. The realization step of trustworthy AI involves seven key requirements that AI systems should implement and meet throughout their entire life cycle using technical or nontechnical methods \cite{busch}. The seven requirements are:
\begin{enumerate}
    \item Human agency and oversight include fundamental human rights, human agency, and human oversight.
    \item Technical robustness and safety include resilience to attack and security, a fallback plan, and general safety, accuracy, reliability, and reproducibility.
    \item Privacy and data governance include respect for privacy, data quality and integrity, and data access.
    \item Transparency includes traceability, explainability, and communication
    \item Diversity, non-discrimination, and fairness include the avoidance of unfair bias, accessibility, and universal design, and stakeholder participation
    \item Societal and environmental wellbeing includes sustainability and environmental friendliness, social impact, society, and democracy
    \item Accountability includes auditability, minimization, and reporting of negative impact, trade-offs, and redress. 
\end{enumerate}

The World Health Organization (WHO) has issued ethical guidance, Ethics \& Governance of AI for Health, on developing and maintaining responsible AI systems, identifying some key AI principles as foundational pillars that guide the use of AI in healthcare in 2021 \cite{who_ethical}.
\begin{enumerate}
    \item Human autonomy: AI systems must respect and support individuals' rights to make informed decisions about their own health and care.
    \item Human well-being, human safety, and the public interest: AI should promote health benefits, avoid harm, and serve the broader public good.
    \item Transparency, explainability, and intelligibility: AI processes and decisions should be open, understandable, and interpretable by users.
    \item Responsibility and accountability: Developers and deployers of AI must be answerable for its outcomes and ensure mechanisms for oversight and redress.
    \item Inclusiveness and equity: AI should be designed and implemented to be accessible, fair, and to reduce health disparities across populations.
    \item Responsive and sustainable AI: AI systems must adapt to changing contexts, be sustainable over time, and minimize environmental impact.
\end{enumerate}

As concerns about the ethical and social impact of AI continue to grow, many developed countries have started creating national AI ethics frameworks. These frameworks aim to ensure that AI technologies are designed and used responsibly, protecting human rights, privacy, and fairness. Table \ref{tab:ethical-summary} presents a comparative summary of the key ethical principles across different organizations and countries.


\begin{table}[t]
\centering
\caption{Summary of major ethical frameworks from different organizations and countries}
\label{tab:ethical-summary}
\renewcommand{\arraystretch}{1.2}
\begin{tabular}{l|c|c|c|c|c|c|c|c|c}
\hline
\textbf{Ethical Domain} & \textbf{Belmont} & \textbf{DoH} & \textbf{Asilomar} & \textbf{IEEE EAD} & \textbf{OECD} & \textbf{WHO} & \textbf{EU} & \textbf{Australia} & \textbf{UK} \\
\hline
Autonomy         & \tick &       &        &        &        & \tick  &        &        &        \\
Beneficence      & \tick &       &        &        &        &        &        &        &        \\
Nonmaleficence   & \tick &       &        &        &        &        &        &        &        \\
Justice          & \tick &       &        &        &        &        &        &        &        \\
Human values     &       &       & \tick  & \tick  & \tick  &        & \tick  & \tick  &        \\
Accountability   &       &       &        & \tick  & \tick  & \tick  & \tick  & \tick  & \tick  \\
Fairness         &       &       &        &        &        &        & \tick  & \tick  & \tick  \\
Transparency     &       &       & \tick  & \tick  & \tick  & \tick  & \tick  & \tick  & \tick  \\
Well-being       &       & \tick &        & \tick  & \tick  & \tick  & \tick  & \tick  &        \\
Privacy          &       &       & \tick  &        &        &        & \tick  & \tick  &        \\
Explainability   &       &       &        &        & \tick  &        &        &        &        \\
Safety           &       &       & \tick  &        & \tick  &        & \tick  & \tick  &        \\
Responsibility   &       &       & \tick  &        &        & \tick  &        &        &        \\
Equity           &       & \tick &        &        &        & \tick  &        &        &        \\
Sustainability   &       &       &        &        &        & \tick  &        &        &        \\
Contestability   &       &       &        &        &        &        &        & \tick  &        \\
Informed Consent &       & \tick &        &        &        &        &        &        &        \\
Community Engage &       & \tick &        &        &        &        &        &        &        \\
Scienctific Base &       & \tick &        &        &        &        &        &        &        \\
Lawfulness       &       &       & \tick  &        &        &        &        &        &        \\
Human Control    &       &       & \tick  &        &        &        &        &        &        \\
\hline
\end{tabular}
\end{table}

In 2020, the United Kingdom strengthened its commitment to responsible data use through the Data Ethics Framework, which provided guidance on transparency, accountability, and fairness in AI and data-driven technologies, ensuring privacy, security, and public trust remained at the core of all data practices \cite{uk2020}. It includes:
\begin{enumerate}
    \item Transparency: Organizations should be open about how and why data is collected, used, and shared.
    \item Accountability: Individuals and institutions using data must take responsibility for their decisions and the impact of their systems.
    \item Fairness: Data and AI systems should be designed and used to treat people equally and avoid discrimination.
\end{enumerate}

Australia has developed a set of eight AI Ethics Principles designed to guide responsible AI development and deployment in 2021 \cite{aus_8ai}. These principles are intended to complement existing regulations and practices of Australia. They emphasise human rights, fairness, accountability, and safety to ensure AI benefits individuals, communities, and the environment. 
\begin{enumerate}
    \item Human, societal, and environmental well-being: AI should deliver benefits for individuals, society, and the environment, aligning with sustainable and equitable development, and must avoid harm.
    \item Human-centred values: AI must respect human rights, diversity, and autonomy, avoiding manipulation and deception, including human dignity, freedom of choice, cultural diversity, and non-coercion.
    \item Fairness: AI should be inclusive, accessible, and free from discrimination to ensure equitable treatment and not reinforce bias.
    \item Privacy protection and security: AI should uphold privacy rights and data protection with strong governance of how data is collected, stored, and used.
    \item Reliability and safety: AI must reliably operate according to its intended purpose with robust risk management, and it should be accurate, reproducible, and proportionate in safety measures.
    \item Transparency and explainability: AI processes and decisions should be explainable and disclosed responsibly, and users should understand when and how AI is influencing them.
    \item Contestability: When AI significantly impacts people, communities, or environments, outcomes should be open to challenge, and redressing mechanisms must be accessible, timely, and effective
    \item Accountability: Those responsible for AI design, deployment, and operation must be identifiable and answerable.
\end{enumerate}

The U.S. does not have a single, unified AI framework but instead has a combination of laws, executive orders, and guidelines that have been in development for several years. The most recent and significant actions were taken in 2025 by the White House Administration \cite{us2025}.

\subsection{Ethical Considerations in Mental Health/Related Work}

AI-powered mental health apps bring several unique ethical challenges, such as privacy risks, bias in algorithms, and the potential for psychological harm. These apps often collect very sensitive personal data, like users’ mental health history, mood patterns, and behavior, which makes it essential to store information safely, get clear user consent, and follow privacy laws to prevent misuse. Moreover, AI models can be unfair because they are trained on data that doesn’t include everyone equally. This means the results might not work well for people from different backgrounds, which can make mental health apps less helpful for the later groups and create gaps in proper care. Also, depending too much on AI chatbots for emotional support can be risky, especially if people think the advice is the same as seeing a real therapist. This is crucial during a crisis, where professional help is really needed. Apparently, AI-based app support is a substitute for professional therapy with automated interactions, which may lead to wrong prescriptions. Misinterpretation of AI-generated recommendations or chatbot responses could also negatively impact users.

Haque and Rubya provide an exploratory observation of 6245 user reviews of 10 mental health mobile chatbots (ADA, Chai, Elomia, Mindspa, Nuna, Serenity, Stresscoach, Woebot, Wysa, Youper) available in the Google Play Store and Apple App Store. They reviewed app descriptions and inspected key information of the apps using observation notes based on six main themes: purpose of the app, targeted concerns, conversation style, the types of media the chatbot uses, crisis support, and what evidence-based techniques it follows \cite{haque2023}. This analysis of user reviews helps understand common patterns in the text while keeping the context of user experiences in mind. They found that chatbots serve human-like interactions with a personalized feel, which users positively receive. In contrast, improper and pre-registered responses and assumptions about the personalities of users led to a loss of interest. Though mobile chatbots are always accessible and convenient, users can become overly attached to them and prefer app use over interacting with friends and family. This may lead to non-maleficence and risk intensification. They claimed that chatbots used in the study created a safe, judgment-free space where users felt more comfortable sharing personal and sensitive information. However, relying too much on technology can be risky, as it may lead to feelings of isolation or fail to provide enough help during a real crisis.

Mental health apps raise issues due to a limited focus on emotional impacts and human relationships in this domain. The \textit{Ethics of Care} approach, proposed by \cite{ethicare}, encompasses the principles that can be applied in practice: (i) mapping the relationships in the process of AI development and implementation which includes the developers, the medical team, the user or the patient, and his or her family; (ii) caring and being responsible for others to adopt certain responsibilities toward patients in mental health considering the specific circumstances and context; (iii) questioning social structures constructing relationships to adhere to similar duties as those for therapists when acting in mental health; and (iv) accepting and reinforcing emotions. This framework addresses the gaps for relational ethics, emotional well-being, and the broader social implications of AI. Nevertheless, challenges remain in this nonbinding principle and the lack of enforcement mechanisms.  

Saeidnia et al. conducted a systematic review to explore the ethical issues of using artificial intelligence (AI) interventions in mental health and well-being \cite{saeidnia2024}. They reviewed 51 research papers published between 2014 and 2024 that focused on ethics in mental health. The study aimed to identify the main ethical concerns, understand how ethical principles can be applied, and highlight best practices and guidelines for using AI responsibly in mental health care. The researchers identified 18 key ethical concerns, including privacy and confidentiality, informed consent, bias and fairness, transparency and accountability, autonomy and human agency, and safety and efficacy. They also suggested several recommendations to help address these growing ethical challenges.

Chaudhry et al. analyzed 159 user reviews of the Wysa mental health app and identified seven key themes: (i) a trusting environment fosters well-being, (ii) ubiquitous access provides real-time support, (iii) AI limitations negatively impact the user experience, (iv) perceived effectiveness of Wysa, (v) the desire for cohesive and predictable interactions, (vi) the appreciation of humanness in AI, and (vii) the need for improvements in the user interface \cite{chaudhry2024}. The study highlights both the strengths and weaknesses of AI-based mental health chatbots and provides suggestions for improving user experience through targeted design updates.

A scoping review of 101 articles published in 2018 or later identified 10 ethical themes in conversational AI-based mental health apps: (i) safety and harm; (ii) explicability, transparency, and trust; (iii) responsibility and accountability; (iv) empathy and humanness; (v) justice; (vi) anthropomorphization and deception; (vii) autonomy; (viii) effectiveness; (ix) privacy and confidentiality; and (x) concerns for health care workers’ jobs \cite{mehrdad2025}.

\subsection{Need for an Empirical Investigation of Ethical Concerns}
Existing ethical frameworks provide high-level theoretical principles that may fail in assessing real-world compliance found in spontaneous app user reviews directly from users. To the best of our knowledge, no research has systematically analyzed user experiences to investigate ethical concerns. Emerging user complaints and case studies suggest that existing pre-defined frameworks may not fully cover some ethical concerns. This leaves a persistent gap between theoretical ethical commitments and their practical implementation in evaluating user perspectives. Emerging evidence from user complaints, litigation, and case studies underscores potential misalignments between existing ethical guidelines and the challenges encountered in practice. For instance, the algorithmic training of an AI-based app behaves differently for vulnerable populations. These discrepancies highlight a need for user evidence-based evaluation of ethical guidelines to ensure their relevance and responsiveness to the real world.

This study aims to bridge this gap by analyzing about 66K user reviews of AI-based mental health applications to empirically assess which ethical concerns are most frequently mentioned by users. Through natural language processing (NLP) techniques, we systematically categorize user-expressed ethical concerns and compare them with those outlined in existing ethical frameworks. The methodology involves (i) scraping and preprocessing textual data from app reviews, (ii) applying topic modeling to detect recurring ethical themes, (iii) mapping these topics to the ethical principles enumerated in the data (from user perspective), and (iv) assessing the sentiment of the reviews, whether the reviews are supporting or rejecting the corresponding ethical principles. This approach will provide new insights into whether current AI ethical guidelines sufficiently address real-world ethical challenges in mental health AI applications.

\section{Method}
\label{method}
In this section, we describe the methodology of this work, which begins with the selection of mobile mental health chatbots for user reviews. Further, we extend the design of a pipeline for identifying and evaluating ethical considerations from the reviews, adopting an NLP-based approach to analyze user-generated app reviews in order to examine how effectively the user app reviews support or reject any ethical principles from the perspective of end-users. 


\subsection{Selection of Apps}

This study was reviewed as not a human subjects research by the Institutional Review Board (IRB) of our home institution. In this empirical study, we considered five mental health apps: Wysa, Woebot, Youper, Sintelly, and Elomia. All four apps are in English and freely available on Google Play and Apple App Store. The common features include an AI-powered chatbot-based personalized mental health assistant that helps users understand and manage their emotions for self-improvement and support. They combine cognitive behavioral therapy (CBT) techniques, meditations, and journaling exercises. They act like a friendly, anonymous conversation partner for emotional support, mood tracking, self-reflexive exercises, and lighthearted interactions to support mental well-being through short, guided conversations. We chose chatbots like apps because users get more engaged with these apps through conversation, which makes them suitable for analyzing ethical considerations. This human-apps interactive nature of these experiences provides valuable data for analyzing how ethical principles are perceived, accepted, or challenged in real-world AI systems. Table \ref{tab:apps-overview} presents an overview of the apps. We obtained 88,241 reviews from Google Play Store and Apple App Store, where users voluntarily posted their feedback on the specific app pages after using the app. Google Play Store ensures the reviews are verified and written after installing the app. Similarly, Apple App Store also ensures the reviews are verified.

\begin{table}[t]
\centering
\caption{Overview of the apps reviewed in this study}
\label{tab:apps-overview}
\renewcommand{\arraystretch}{1.2}
\begin{tabular}{p{0.25\textwidth}|c|c|c|c}
\hline
\textbf{App Name} & \textbf{Overall ratings} & \textbf{No. of reviews} &
\textbf{No. of downloads} & \textbf{Released in} \\
\hline
\multirow{2}{4cm}{Wysa: Anxiety, therapy chatbot \cite{wysaplay,wysaapp}} 
  & In Play Store: 4.6 & 151K & 4,459,977 & 2016 \\ 
  & In App Store: 4.8 & 24K & Unknown & Unknown \\ \hline

\multirow{2}{4cm}{Woebot: The Mental Health Ally \cite{woeplay,woeapp}} 
  & In Play Store: 3.5 & 12.5K & 607,437 & 2018 \\ 
  & In App Store: 4.6 & 6.2K & Unknown & Unknown \\ \hline

\multirow{2}{4cm}{Youper -- CBT Therapy Chatbot \cite{youplay,youapp}} 
  & In Play Store: 3.9 & 49.5K & 1,477,882 & 2015 \\ 
  & In App Store: 4.8 & 15.4K & Unknown & Unknown \\ \hline

\multirow{2}{4cm}{Sintelly: CBT Therapy Chatbot \cite{sinplay,sinapp}} 
  & In Play Store: 4.3 & 10.3K & 1,186,433 & 2017 \\ 
  & In App Store: 4.6 & 308 & Unknown & Unknown \\ \hline

\multirow{2}{4cm}{Elomia: Mental Health AI \cite{eloplay,eloapp}} 
  & In Play Store: Unknown & 6 & $\geq$ 500 & 2025 \\ 
  & In App Store: 4.5 & 1.1K & Unknown & Unknown \\ \hline
\end{tabular}
\end{table}

\subsection{Data Collection}

The Google Play Store reviews are scraped using Google Play Scraper \cite{scraper}, a Python-based API library that allows crawling the app web pages. For Apple App Store, user reviews were retrieved using their official iTunes Search API and RSS feed services \cite{apple_itunes_api_2024}. After scraping, we carefully go through all the reviews for human perceptions of the reviews. We excluded some reviews considering some exclusion criteria, such as a minimum word length of five words, and excluding other languages except English. This results in a final set of 65,948 reviews for analysis. Google Play Store user reviews are publicly accessible, requiring no gatekeeper for access. According to Google Play policy \cite{playrev,playpolicy}, publicly posted reviews are visible to all Play Store users and may be utilized by developers to gain insights into user experiences. The App Store has provided the same guidelines \cite{apppolicy}. To safeguard the reviewers' privacy, we adjusted the wording of original quotes slightly in this article while preserving their original meaning. This approach aligns with the recommendations of Eysenbach and Till \cite{eysenbach2001}, who suggest modifying the content of publicly posted user reviews to reduce the likelihood of the original quotes being located through search engine queries. Additionally, we did not collect any identifiable information, such as names or email addresses, that the reviewers may have posted on the Google Pay page to ensure reviewers’ anonymity. The number of reviews from the Apple App Store is lower because data were collected using Apple’s official RSS feed, which provides only a limited set of the most recent reviews per app and does not support full historical retrieval.

\section{Analysis}

\begin{figure}[b]
\centering
\includegraphics[width=\linewidth]{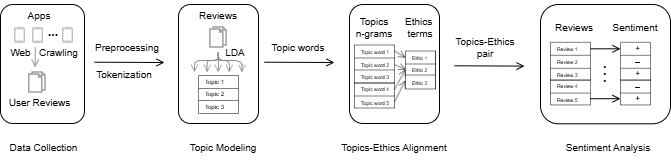}
\caption{Overview of proposed framework}
\label{fig:method}
\end{figure}

We introduce a four-layer NLP-driven framework for the task of evaluating ethical considerations: (i) preprocessing stage, (ii) topic modeling stage, (iii) topic-ethics alignment stage, and (iv) sentiment analysis stage. Figure \ref{fig:method} illustrates an overview of the proposed framework. The preprocessing stage will scrape raw data, examine readability tests of reviews, and perform basic preprocessing to fit into the model. Relevant topics will be extracted using the LDA model in the extraction stage. Finally, in the alignment stage, the extracted topics are aligned with pre-defined ethical concerns, and the polarity of the reviews is found using aspect-based sentiment analysis techniques.

\subsection{Preprocessing Stage}
\begin{figure}[b]
    \centering
    \includegraphics[width=0.8\linewidth]{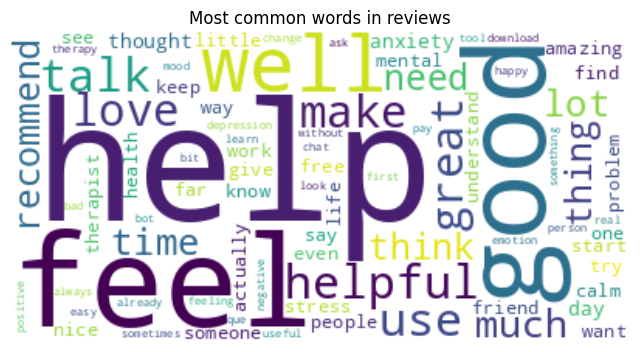}
    \caption{Word cloud of topic words}
    \label{fig:top_words}
\end{figure}

\begin{figure}[t]
    \centering
    \begin{subfigure}[b]{0.48\textwidth}
        \centering
        \includegraphics[width=\textwidth]{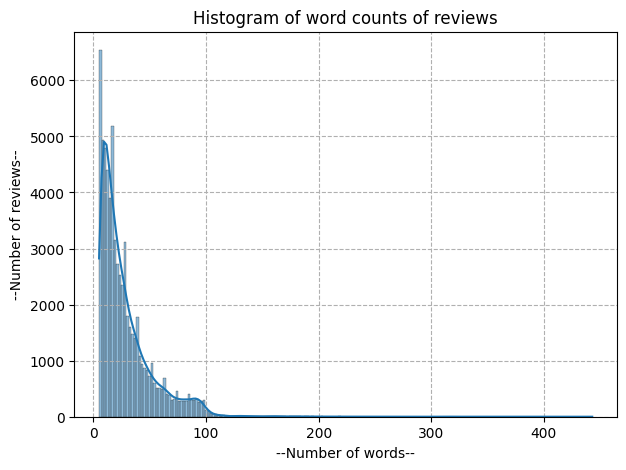}
        \caption{Words}
    \end{subfigure}
    \hfill
    \begin{subfigure}[b]{0.48\textwidth}
        \centering
        \includegraphics[width=\textwidth]{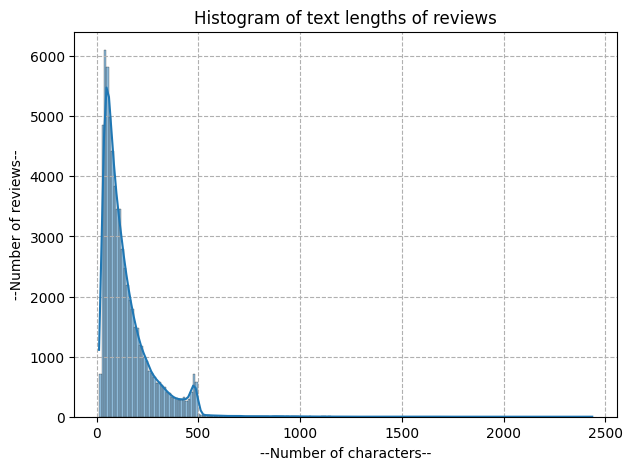}
        \caption{Characters}
    \end{subfigure}
    \caption{Distribution of words and characters in the corpus}
    \label{fig:distribution}
\end{figure}

The first stage of our study focuses on the construction of a comprehensive, large-scale data set derived from user-generated content. We collected a total of 88,241 user reviews from the Google Play Store and Apple App Store using web scraping techniques. These reviews were aggregated app-wise, capturing metadata such as review text and publication dates. After collecting raw data, we observed poor linguistic features of the user-generated texts. User reviews are characterized by informal language, grammatical inconsistencies, emojis, abbreviations/short-form words (y'all=you all, rlly=really, tbh= to be honest), and typo errors. Non-English texts, English stopwords, and domain-related words have also been removed. So, a rigorous preprocessing pipeline was implemented, aiming to enhance the readability of the texts. We used several preprocessing techniques, such as removal of non-alphanumeric characters, punctuation marks, dates, emojis, redundant whitespace, HTML tags, and URLs using regular expressions. The reviews with exceptionally low readability scores were discarded to ensure data quality. Lemmas consisting of fewer than three characters were excluded from the dataset. A text-based normalization experiment was conducted to ensure robustness against the noise inherent in free-text user inputs. A total of 65,948 remain. The statistical properties of the curated dataset are summarized in Table \ref{tab:review_stats} while the top occurring words are shown in Figure \ref{fig:top_words}. The distributions of words and characters are shown in Figure \ref{fig:distribution}. 

\begin{table}[t]
\centering
\caption{Statistics of user review texts}
\label{tab:review_stats}
\renewcommand{\arraystretch}{1.2}
\begin{tabular}{l|c|c|c}
\hline
\textbf{Statistics} & \textbf{Play Store} & \textbf{App Store} & \textbf{Total} \\ \hline
No. of reviews & 62,861 & 3,087 & 65,948 \\ 
Total sentences & 166,074 & 10,056 & 176,130 \\ 
Total words & 1,737,132 & 146,715 & 1,883,847 \\ 
Total characters & 8,809,545 & 743,870 & 9,553,415 \\ \hline
Max. sentences per review & 47 & 29 & 47 \\ 
Min. sentences per review & 1 & 1 & 1 \\ 
Avg. sentences per review & 2.64 & 3.26 & 2.67 \\ \hline
Max. words per review & 421 & 443 & 443 \\ 
Min. words per review & 5 & 5 & 5 \\ 
Avg. words per review & 27.63 & 47.53 & 28.57 \\ \hline
Max. characters per review & 2,241 & 2,435 & 2,435 \\ 
Min. characters per review & 12 & 15 & 12 \\ 
Avg. characters per review & 140.14 & 240.97 & 144.86 \\ \hline
\end{tabular}
\end{table}

\subsection{Topic Modeling Stage}
\begin{figure}[b]
    \centering
    \includegraphics[width=0.7\linewidth]{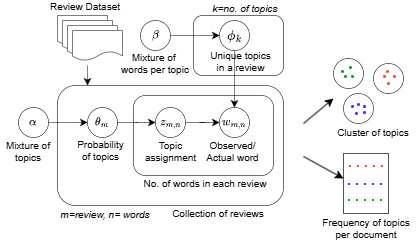}
    \caption{How LDA works in the review dataset. $\alpha$ and $\beta$ are Dirichlet parameters.}
    \label{fig:lda}
\end{figure}

We used the Latent Dirichlet Allocation (LDA) model to uncover hidden topics from user reviews of mobile apps~\cite{blei2003latent}. LDA is an unsupervised, probabilistic learning model that treats each document (in our case, each user review) as a mixture of latent topics, and each topic is represented by a probability distribution of words, indicating how likely each word is to appear within that topic. Our LDA takes two inputs: a corpus and a dictionary. The corpus represents the set of all user reviews and is denoted by $\mathcal{C} = {d_1, d_2, \ldots, d_M}$ and the dictionary $\mathcal{D}$ is a mapping of all unique tokens to integer indices, that is, 
$\mathcal{D} = \{ (w_j, \text{id}_j) \,|\, w_j \in \mathcal{V} \}$, 
where $\mathcal{V}$ denotes the vocabulary. 
Figure \ref{fig:lda} illustrates how the LDA model was applied to the user reviews.

For training, LDA represents each document (review) as a distribution over $K$ topics, and each topic as a distribution over words. A $K$-dimensional Dirichlet random variable $\theta$ defines the topic mixture for document $d$. Let, the hyperparameters $\alpha$ (mixture of topics) and $\beta$ (mixture of words), the joint distribution of the topics $\theta$, a set of latent topics $z$, and a set of observed actual words $w$ is given as:

\begin{equation}
p(\theta,\, z,\, w \,|\, \alpha,\, \beta) 
= p(\theta \,|\, \alpha) 
  \prod_{n=1}^{N} p(z_n \,|\, \theta) p(w_n \,|\, z_n,\, \beta)
\end{equation}

where $p(z_n \,|\, \theta)$ is simply $\theta_i$ for the unique topic $i$ such that $z^i_n = 1$.  
Integrating over $\theta$ and summing over all topic assignments $z$, we obtain the probability of a review document:

\begin{equation}
p(w \,|\, \alpha,\, \beta) 
= \int p(\theta \,|\, \alpha) 
  \left( \prod_{n=1}^{N} 
    \sum_{z_n} p(z_n \,|\, \theta) p(w_n \,|\, z_n,\, \beta)
  \right) d\theta
\end{equation}

Finally, assuming independence across documents, we derive $p(D \,|\, \alpha,\, \beta)$ by multiplying the probabilities $p(w \,|\, \alpha,\, \beta)$ of all documents appear in the corpus. The likelihood of topics for the entire corpus is:

\begin{equation}
p(D \,|\, \alpha,\, \beta) 
= \prod_{d=1}^{M} 
  \int p(\theta_d \,|\, \alpha) 
  \left( \prod_{n=1}^{N} 
    \sum_{z_{dn}} p(z_{dn} \,|\, \theta_d) p(w_{dn} \,|\, z_{dn},\, \beta)
  \right) \, d\theta_d
\end{equation}


We used the \textit{Gensim} Python library to implement the LDA algorithm~\cite{rehurek2011gensim}. The input corpus was built as a term–document matrix from the cleaned user reviews. Each entry in the matrix represents the frequency of a word or phrase in a given review. This allows the model to learn the co-occurrence patterns of latent ethical themes.

The quality of the model depends on several hyperparameters, includes the number of topics ($K$), the number of training passes ($passes$), the batch size ($chunk\_size$), etc. The optimal number of topics is determined empirically by evaluating \textit{Coherence score}. Coherence score assesses the semantic interpretability of topics combining a sliding-window co-occurrence using Normalized Pointwise Mutual Information (NPMI) approach and cosine similarity of context vectors~\cite{roder2015exploring}. To illustrate, the coherence score of a topic \( T \) with top words \( \{w_1, w_2, \ldots, w_M\} \) is computed using the vector-based coherence metric \( C_v \). It is defined as:

\begin{equation}
C_v(T) = \frac{1}{|S|} \sum_{(w_i, w_j) \in S} 
\cos\left( \text{NPMI}(w_i, \cdot), \, \text{NPMI}(w_j, \cdot) \right)
\end{equation}
while, 
\begin{equation}
\text{NPMI}(w_i, w_j) = 
\frac{\log \frac{P(w_i, w_j)}{P(w_i)P(w_j)}}{-\log P(w_i, w_j)}
\end{equation}
where \( S \) refers to the set of all word pairs within a topic (\( S=20 \), in our case); \(\text{NPMI}(w_i, \cdot)\) represents the vector of NPMI scores between the word \( w_i \) and all other words in the same topic; and \( P(w_i, w_j) \) denotes the probability that the two words appear together within a sliding window across the corpus.

\begin{figure}[b]
    \centering
    \includegraphics[width=0.6\linewidth]{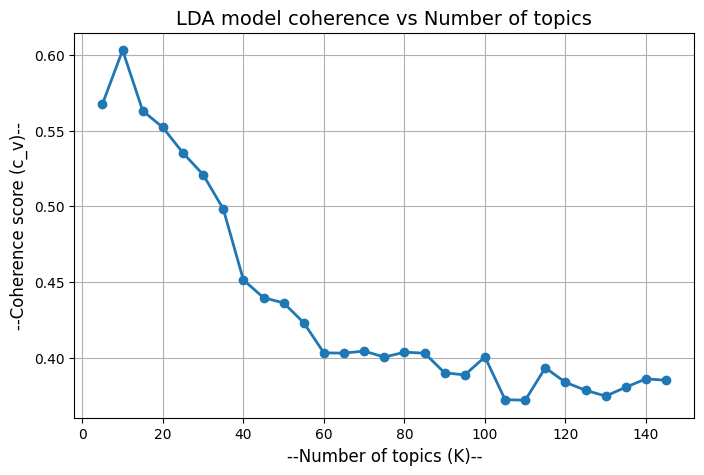}
    \caption{Coherence score trend}
    \label{fig:coherence_curve}
\end{figure}

The cosine similarity measures how closely related the NPMI vectors are in meaning, and the average of these values gives the overall topic coherence. In simple terms, $C_v$ produces higher scores when the top words in a topic often appear together in similar contexts. The score ranges from 0 to 1, where a higher value means the topic is more meaningful and easier to interpret. To get the optimal number of topics ($K$), we trained several LDA models with different $K$ values ranging from 5 to 150.  The coherence scores for different $K$ values are shown in Fig.~\ref{fig:coherence_curve}. The highest coherence (0.603) was observed at $K = 10$, after which the values dropped sharply around $K \approx 35$, suggesting the topics became less consistent. So, $K = 35$ was chosen as the upper limit, and the top 20 words from each topic were used for analysis.

After training, the model generates a topic distribution for each document, denoted as $\theta_d = { \theta_{d,1}, \ldots, \theta_{d,K} }$, and a word distribution for each topic, denoted as $\phi_k = { \phi_{k,1}, \ldots, \phi_{k,V} }$. Each topic $k$ gives us the most frequent and relevant words, listed in order of their likelihood. These top words capture the main ideas expressed by users and are later used in the topic–ethics alignment stage.

\subsection{Topic-Ethics Alignment Stage}
\begin{figure}[bp]
    \centering
    \includegraphics[width=0.7\linewidth]{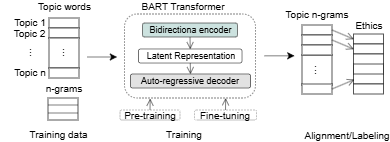}
    \caption{Alignment stage}
    \label{fig:alignment}
\end{figure}

The third stage of our framework aims to align the extracted topics from user reviews with existing ethical themes and also identify new ethical themes which are not covered by the existing frameworks. As shown in Figure \ref{fig:alignment}, the process follows an alignment mechanism.

To detect the ethical themes, we perform a semantic alignment between the topics discovered through LDA and a curated set of ethical dimensions derived from the existing frameworks. The predefined ethical taxonomy, denoted as: \(\mathcal{E}_{\text{known}} = \{e_1, e_2, \ldots, e_P\},\) includes existing ethical principles such as privacy, beneficence, non-maleficence, autonomy, etc. In addition, the model is also intended to discover new ethical considerations: \(\mathcal{E} = \mathcal{E}_{\text{known}} \cup \mathcal{E}_{\text{new}}.\)

For semantical alignment, each topic $\tau_j$ obtained from the LDA model is represented as a sequence of its most probable words (or representative terms formed by top-$n$ words). To map each topic $\tau_j$ to its corresponding ethical category, we used the Bidirectional and Auto-Regressive Transformer (BART) model available in the HuggingFace library~\cite{yin2019benchmarking,bart2019meta}. BART is a transformer-based zero-shot classification model which is pre-trained for various text understanding tasks, such as text classification, paraphrase detection, and textual inference. It makes it suitable for semantically aligning topics to the associated ethics term ~\cite{lewis2020bart}. We perform fine-tuning of the model on a manually curated dataset of ethical aspects.

Formally, given a topic $\tau_j$ and an ethical taxonomy $\mathcal{E}_{\text{known}}$, the alignment function is defined as:
\[
f_{\text{align}} : \mathcal{T} \rightarrow \mathcal{E}, \qquad
\hat{e}_j = f_{\text{align}}(\tau_j) = \arg\max_{e_p \in \mathcal{E}} \, \text{Sim}(\tau_j, e_p)
\]
where $\text{Sim}(\tau_j, e_p)$ denotes the semantic similarity score between the topic description and the textual definition of the ethical principle $e_p$.  
The similarity is computed using contextual embeddings from the BART encoder, expressed as:
\[
\text{Sim}(\tau_j, e_p) = \cos(\mathbf{v}_{\tau_j}, \mathbf{v}_{e_p})
\]
where $\mathbf{v}_{\tau_j}$ and $\mathbf{v}_{e_p}$ are high-dimensional embeddings of the topic and ethical themes, respectively. 

If the value of $\text{Sim}(\tau_j, e_p)$ exceeds a threshold (we set 0.5), the topic is assigned to the corresponding ethical principle $e_p \in \mathcal{E}_{\text{known}}$; otherwise, it is marked as a potential novel ethical consideration, denoted as $\hat{e}_j \in \mathcal{E}_{\text{new}}$. These newly found novel ethical considerations are later validated through a human-interpreted qualitative review to ensure conceptual coherence. 


\subsection{Sentiment Analysis Stage}
The final stage of our framework determined the sentiment of user reviews whether a review supports or opposes a particular ethical aspect. This step helps identify whether users express a positive or negative attitude toward each ethical dimension. In this case, we didn't consider the numeric ratings provided by the user while writing a review. Because, numeric ratings donot represent any contextual aspects specially related to ethics.

Sentiment analysis, also known as opinion mining, involves studying emotions, opinions, and subjective expressions in text~\cite{liu2012sentiment}. In the context of mobile health applications, user reviews are expressed spontaneous emotional reactions and personal experiences after using the apps. By analyzing these expressions, we can gain an understanding of how users perceive the ethical reliability and overall usability of these apps.

Let $\mathcal{R} = \{r_1, r_2, \ldots, r_N\}$ be the set of all preprocessed user reviews. The objective is to estimate the polarity of sentiment $s_i$ for each review, where
\( s_i = g(r_i) \in \{-1, +1\} \) represents the positive and negative sentiment classes, which means that they support or reject the corresponding ethics, respectively. Note that each review $r_i$ may be associated with one or more ethical labels $E_i = \{e_1, e_2, \ldots, e_m\}$ retrieved from the alignment stage.

To obtain polarity estimation, we implement an aspect-based sentiment analysis (ABSA) model~\cite{hu2004mining,pontiki2016semeval}, in which the sentiment is computed with respect to the aspect (in our case, ethical themes) expressed in the review. We already aligned the ethical aspects to the reviews in the previous stage. The sentiment probability distribution for a given review $r_i$ and ethical aspect $e_j$ is expressed as:
\[
p(s \,|\, r_i, e_j) = \text{softmax}(f_{\text{sent}}([h_{r_i}; h_{e_j}]))
\]
where $h_{r_i}$ and $h_{e_j}$ denote the contextual embeddings of the review and ethical aspect, respectively and $f_{\text{sent}}(\cdot)$ is a neural classification function implemented using a transformer encoder RoBERTa~\cite{liu2019roberta}.  
The predicted sentiment class $\hat{s}_{i,j}$ is obtained as:
\[
\hat{s}_{i,j} = \arg\max_{s \in \{-1, +1\}} p(s \,|\, r_i, e_j)
\]

\section{Results}

Our framework aims to explore users' concerns about the ethical issues in AI-based mental health chatbots. To achieve this, the system first generates topic words from collected user reviews, aligns these topics with ethical themes, and finally measures the sentiment behind those reviews.

In the topic modeling stage, we identified common patterns and recurring ideas from thousands of user reviews. These patterns include topic words that capture user experiences. Table \ref{tab:lda_topics} shows some examples of the topics found by the LDA model. Each topic groups together words with related meanings. We identified 457 single words, 163 two-word phrases, and 80 three-word phrases across all topics.

\begin{table}[t]
\centering
\renewcommand{\arraystretch}{1.2}
\caption{Top words extracted from LDA model (Topic 26-35)}
\label{tab:lda_topics}
\begin{tabular}{c|p{0.72\textwidth}|c}
\hline
\textbf{Topic \#} & \textbf{Top words (Representative $n$-grams)} & \textbf{No. of reviews} \\
\hline
26 & talk, someone, feel, like, friend, need, help, always, feeling, listen, great, share & 6889 \\
27 & recommend, help, highly, everyone, die, great, think, try, hard, time, struggle & 2415 \\
28 & good, self, pretty, far, care, personal, tracker, help, see, pocket, need, try & 2285 \\
29 & people, life, grateful, need, help, save, disappointed, want, guess, trigger, fine & 635 \\
30 & mood, track, experience, situation, bad, improve, keep, help, emotion, explain & 776 \\
31 & use, start, easy, super, support, helpful, great, seem, short, time, impressed & 1864 \\
32 & awesome, application, message, matter, meet, confidence, present, helpful, trauma & 362 \\
33 & give, check, daily, idea, advice, perfect, test, great, therapy, never, try & 874 \\
34 & back, come, tell, everything, point, get, see, stop, show, log, away, time, happen & 774 \\
35 & help, anxiety, depression, stress, deal, attack, social, suffer, panic, relationship & 3641 \\
\hline
\end{tabular}
\end{table}



In the next stage, we align the extracted topics to the existing and new ethical concepts based on the similarities of the words. Topics~10, 20, 23, and 30, are grouped by words such as \textit{help, improve, progress, struggle, cope, keep}, represent a semantic relationship with the principle of \textit{beneficence} and \textit{compassion}. Users often describe relief and emotional encouragement from a chatbot as a caring companion. Topics~5, 9, 11, 16, and~32 reference words \textit{anxiety, stress, sleep, relax, cope}, which highlight issues related to \textit{non-maleficence} and the emergent theme of \textit{emotional dependency}. 

Topics~1, 2, 17, and~28 align with usability aspects of technology, connecting terms such as \textit{use, access, sign, interface, session}. The words from Topics~14, 16, and~24 like \textit{thought, positive, change, turn, understand} indicate the change in user perspective which aligns the ethical theme like, \textit{human well-being} and \textit{cultural sensitivity}. Table~\ref{tab:ethics-alignment} shows more mapping results between topic clusters, representative $n$-grams, and corresponding ethical considerations. Table \ref{tab:ethics} portrays all ethical considerations found in the reviews. 

\begin{table}[tbp]
\centering
\caption{Top 15 topics and their aligned ethical aspects with reasoning}
\label{tab:ethics-alignment}
\renewcommand{\arraystretch}{1.25}
\begin{tabular}{c|p{3.2cm}|p{3.5cm}|p{6.5cm}}
\hline
\textbf{Topic \#} & \textbf{Representative terms} & \textbf{Aligned ethical aspect} & \textbf{Reasoning (Human interpretation)} \\
\hline
1 & nice, honestly, aware, rating, reason & Usability, Empathy & Users question if the chatbot is real or human, implying the need for emotional empathy \\ \hline
6 & chat, bot, response, friendly, text, user & Emotional Dependency, Parasocial Attachment & The perception of friendship and continuous conversation suggests potential over-reliance on AI companionship \\ \hline
8 & thank, wysa, developer, appreciate, team, create & Fairness, Equity & Paid or premium features raise fairness concerns around accessibility and equitable service \\ \hline
9 & find, wonderful, positive, comfort, concept, often & Cultural sensitivity, Algorithmix bias & Language barrier represents inclusivity for users with diverse use; Chat tone reflects bias towards specific group \\ \hline
14 & free, pay, subscription, access, feature, worth, money & Human well-being & Conversation with chatbot help users to change their minds into positive emotion and beliefs \\ \hline
15 & feel, help, happy, understand, alone, comfortable & Beneficence, Cognitive overload & User sometimes losing attention during long conversations to make the app understand the problem \\ \hline
17 & personality, relax, interactive, explore, supportive, believe & Clinical oversight & Apps are not clinically validated and asking about the fact misleads users and no clear evidance \\ \hline
19 & download, use, hope, today, continue, year & Accountability & Expectation for responsible handling of user complaints and transparent algorithmic decision hampers \\ \hline
31 & awesome, application, step, message, confidence, trauma & Harm Mitigation & References to panic and attack relate to emotional safety and responsible handling of mental distress \\ \hline
34 & anxiety, depression, stress, panic, attack, relationship & Explainability, Transparency & Chatbot reasoning are not explained clearly and transparency is not assured in conversation \\ \hline
\end{tabular}
\end{table}

\begin{table}[htbp]
\centering
\caption{Existing and Emergent Ethical Considerations in AI-based Mental Health Apps}
\label{tab:ethics}
\renewcommand{\arraystretch}{1.3}
\begin{tabular}{p{0.45\textwidth}|p{0.45\textwidth}}
\hline
\textbf{Existing ethical principles} & \textbf{Emergent ethical concerns} \\ \hline
\begin{enumerate}[topsep=-3pt]
\item Human-centered value / Respect / Empathy
\item Transparency
\item Justice / Equity / Bias minimization / Fairness / Inclusivity
\item Non-maleficence / Harm reduction / Avoidance of harm
\item Beneficence
\item Autonomy / Consent
\item Accountability / Responsibility
\item Sustainability
\item Privacy / Data protection / Confidentiality
\item Safety
\item Accessibility
\setcounter{saveenum}{\value{enumi}}
\end{enumerate}
&
\begin{enumerate}[topsep=-3pt]
\setcounter{enumi}{\value{saveenum}} 
\item Clinical oversight
\item Social well-being
\item Lawfulness
\item Cultural sensitivity
\item Trust / Trust calibration
\item User control
\item Explainability
\item Usability / Responsiveness
\item Emotional dependency
\item Algorithmic bias
\item Digital well-being / Fatigue / Cognitive overload
\end{enumerate}
\\
\hline
\end{tabular}
\end{table}


In the sentiment analysis part of our study, we measured the polarity of each ethic found in the previous stage, as to how users view the topics. 
To do that, we calculated the sentiment of each review. Then, the average sentiment score for each topic, called $S_i$, was calculated by simply adding up the sentiment values of all reviews in that topic and dividing by the number of reviews:
\[
S_i = \frac{1}{M_i} \sum_{j=1}^{M_i} \text{Polarity}(r_j)
\]
where, $\text{Polarity}(r_j)$ means how positive or negative a single review $r_j$ was in topic $i$.

We achieved an F1-score of 0.86 tested on 13,200 reviews (which is 20\% of dataset). Topics about \textit{beneficence}, \textit{autonomy}, and \textit{fairness} had mostly positive feelings ($S_i > +0.5$), while topics about \textit{privacy}, \textit{transparency}, and \textit{dependency} had more negative feelings ($S_i < -0.3$). Table~\ref{tab:ethics-frequency} shows the different ethical categories, how often they appeared, and the average sentiment for each one.

\begin{table}[t]
\centering
\caption{Identified ethical aspects, their frequency of occurrence, and average sentiment polarity.}
\label{tab:ethics-frequency}
\renewcommand{\arraystretch}{1.25}
\begin{tabular}{p{0.48\textwidth}|c|c}
\hline
Ethical Aspect & Frequency (\% of reviews) & Average sentiment score \\
\hline
Human-centered Value/Respect/Empathy & 9.8\% & +0.67 \\
Transparency & 12.5\% & -0.41 \\
Justice/Equity/Bias/Fairness/Inclusivity & 9.1\% & +0.08 \\
Non-maleficence/Harm Reduction/Avoidance of Harm & 6.7\% & -0.12 \\
Beneficence & 21.8\% & +0.73 \\
Autonomy/Consent & 10.6\% & +0.32 \\
Accountability/Responsibility & 7.9\% & +0.14 \\
Sustainability & 4.6\% & +0.11 \\
Privacy/Data Protection/Confidentiality & 17.2\% & -0.54 \\
Safety & 5.3\% & -0.09 \\
Accessibility & 6.4\% & +0.21 \\
\hline
Clinical Oversight (Emergent) & 4.8\% & -0.16 \\
Social Well-being (Emergent) & 7.1\% & +0.29 \\
Lawfulness (Emergent) & 3.9\% & -0.20 \\
Cultural Sensitivity (Emergent) & 5.9\% & -0.29 \\
Trust/Trust Calibration (Emergent) & 8.2\% & +0.25 \\
User Control (Emergent) & 6.8\% & +0.18 \\
Explainability (Emergent) & 5.7\% & -0.33 \\
Usability/Responsiveness (Emergent) & 8.9\% & +0.35 \\
Emotional Dependency/Compassion (Emergent) & 8.3\% & -0.37 \\
Algorithmic Bias (Emergent) & 5.9\% & -0.29 \\
Digital Fatigue/Cognitive Overload (Emergent) & 7.2\% & -0.26 \\
\hline
\end{tabular}
\end{table}

The intergrated topic alignment-sentiment analysis showed a clear pattern towards meaningful interpretation. Positive values like beneficence, fairness, and autonomy were easily aligned, but negative or new issues such as privacy problems, emotional dependency, and algorithmic bias show more diffuse semantic meanings.

To understand these results better, we looked at some topics more closely. We studied the most common words and phrases in each topic to see how people express ethical ideas in their reviews. The findings show that users often share their moral opinions through the emotions in their writing.
Topics that focused on the words like \textit{help, calmness, and progress} (Topics 3, 6, 20, and 33) were aligned with ideas of beneficence, well-being, and human values, and they had strong positive sentiment ($S_i > 0.6$). In contrast, topics about data use, payment limits, or annoying notifications (like Topics 17, 19, 25, and 29) showed negative feelings ($S_i < -0.3$). They were related to privacy, accountability, and newer issues like digital fatigue and cultural sensitivity.

\section{Discussion}

The proposed NLP-driven framework enabled the extraction, alignment, and interpretation of ethical issues embedded within user-generated reviews of AI-based mental health applications. Through sequential stages of topic modeling, ethical alignment, and sentiment aggregation, the framework identified a total of nine distinct ethical considerations, encompassing both recognized and emergent dimensions of AI ethics. 

User reviews of mental health mobile applications reveal ethical concerns that extend well beyond traditional issues of privacy or clinical effectiveness. These reviews provide a grounded, experience-based view of how ethical challenges are encountered in everyday use, often exposing gaps between designers’ intentions and users’ lived realities. Rather than abstract principles, users describe concrete moments of confusion, distress, mistrust, or disappointment, showing how ethics are negotiated through interaction.

\subsection{Beyond Helpfulness: Rethinking Beneficence and Harm}

While many users initially approach mental health apps seeking emotional relief, support, or guidance, reviews suggest that perceived beneficence is fragile and easily undermined. Users often describe situations where advice feels generic, unexplained, or poorly matched to their actual problems. When chatbots ask too many questions, repeat prompts, or fail to grasp context, interactions shift from supportive to emotionally exhausting. These experiences highlight that \textit{doing good} in mental health contexts is not only about providing responses, but about pacing, sensitivity, and situational awareness. Harm, in this sense, is not always dramatic or immediate, it can emerge gradually through emotional fatigue, frustration, or a sense of being misunderstood.

\subsection{Transparency, Explainability, and User Confusion}

A recurring concern across reviews is the lack of clarity about how apps generate suggestions, exercises, or diagnoses. Users want to understand \textit{why} certain messages are shown and \textit{how} personalization works. When explanations are absent or overly technical, trust erodes. This lack of explainability becomes especially problematic in mental health settings, where users may already feel vulnerable or uncertain. Reviews show that unexplained advice can feel intrusive, manipulative, or even dangerous, particularly when it touches on sensitive topics. Transparent communication, therefore, is not just a technical requirement, it is central to emotional safety and informed engagement.

\subsection{Autonomy, Consent, and Control Over Data}

Many users express discomfort with how much data they are asked to share and how little control they feel they have once they do. Reviews frequently mention unclear consent processes, limited options to opt out of data collection, and difficulty deleting past conversations or withdrawing personal information. These experiences challenge the assumption that consent is a one-time action. Instead, users appear to value ongoing autonomy, the ability to pause, leave, or reconfigure their engagement without emotional pressure or hidden consequences. Ethical concerns intensify when users sense that data is silently shared with third parties, especially for advertising or analytics, undermining trust and the perceived integrity of care.

\subsection{Privacy, Security, and Broken Trust}

Mental health data is deeply personal, and reviews reflect heightened sensitivity to privacy failures or vague assurances. Users report anxiety when privacy policies are outdated, frequently changed, or difficult to understand. Others raise concerns about inadequate security measures and misleading promises about confidentiality. These perceptions suggest that privacy is not only about compliance, but about felt safety. When users doubt whether their data is protected, the therapeutic value of the app is fundamentally compromised.

\subsection{Inequity, Accessibility, and Exclusion}

User reviews also surface structural ethical issues related to equity and inclusivity. High subscription costs, aggressive paywalls, and limited free features make access uneven, particularly for users from lower-income backgrounds. At the same time, lack of support for non-English languages, culturally specific expressions, or diverse mental health norms leaves many users feeling excluded. Reviews from non-Western or multilingual users highlight how Western-centric therapy language can feel alienating or irrelevant. Accessibility concerns extend further to users with disabilities, who report that many apps are not designed with visual, auditory, or motor impairments in mind. These issues reveal how ethical design must account for diversity in language, culture, ability, and socioeconomic context.

\subsection{Emotional Dependency and Blurred Boundaries}

An emergent ethical theme in reviews is the risk of emotional over-reliance on AI companions. Some users describe forming strong emotional attachments to chatbots, while others express discomfort with systems that encourage prolonged engagement without clearly acknowledging their limits. When apps fail to communicate that they are not substitutes for human care-or when they simulate companionship too convincingly-users may experience confusion about the app’s role. Tragic real-world cases highlighted in public discourse underscore the potential consequences of blurred boundaries, reinforcing the need for careful calibration between empathy and realism.

\subsection{Accountability and Responsibility in Practice}

Users frequently judge ethical responsibility not by policy statements, but by how developers respond to problems. Reviews mention frustration when harmful outputs, biased responses, or technical failures persist across updates. The absence of clear escalation pathways-especially during moments of crisis-raises serious concerns about accountability. Ethical responsibility, from the user perspective, includes timely updates, visible human oversight, and clear referral mechanisms when risks become critical.

\subsection{Trust as an Ongoing, Fragile Process}

User reviews suggest that trust in mental health apps is not stable. It fluctuates based on interaction quality, transparency, and perceived sincerity. Excessive disclaimers, complex explanations, or inconsistent behavior can overwhelm users rather than reassure them. Trust, therefore, must be carefully calibrated-clear enough to inform, but gentle enough to support emotional well-being. When users feel respected, understood, and safe, trust grows; when they feel confused or manipulated, it quickly collapses.

These findings show that ethical concerns in mental health apps are not isolated issues but interconnected experiences shaped through daily use. User reviews reveal ethics as something lived, negotiated, and felt-emerging at the intersection of technology, emotion, and care. Designing ethical mental health technologies therefore requires moving beyond static principles toward continuous, user-centered ethical reflection grounded in real experiences.

\section{Implications}

Although the findings of this study provide important guidance for improving how AI-based mental health systems are designed and evaluated ethically, it has several implications.

Developers should design AI systems that protect users not only from technical failures but also from emotional harm. Application features such as distress detection, crisis response, and reflective dialogue can help ensure that users are supported safely. Ethically Responsible AI (RAI) should avoid amplifying emotional dependency and be more sensitive for decision-making task and observe user language more closely. They should comply with psychological-safety-by-design framework.

AI-based systems should communicate clearly and compassionately about what they are and what they can do. For example, during a crisis-like conversation, an app might gently remind the user, ``I am an AI assistant, not a human therapist," followed by a reassuring message like, ``I am here to listen and support you." Such emotionally aware communication helps build trust while making sure users understand the limits of AI support.

AI systems build on algorithms where people and not care about. In case of sensitive mental health issue, algorithmic processes should be clear how the decisios are made. Many users are concern about the process reflected in the reviews. Along with the mitigation of algorithmic bias, AI-based apps should use diverse data across different culture. Incorporating local tone of language, cultural- specific metaphors, and emotional expressions can help achieve affective inclusivity. The design of mental health chatbots should respect user autonomy by adopting adaptive interaction pace, periodic reminders to take breaks etc. This can prevent overreliance in AI systems ensuring emotional support without long-term dependency. The developers should update the apps regularly and analyze app reviews which can detect early signals of harm (e.g., dependency or fatigue), and evaluate whether interventions uphold safety.

\section{Conclusion}
In this work we systematically evaluate the ethical considerations of user reviews posted voluntarily in Google Play Store and Apple App Store in five application pages application (e.g., Wysa, Woebot, Youper, Sintelly, or Elomia). We tired to find how individuals experience and interpret AI behavior (how the system behaves vs how users feel about it). This work extend beyond evaluating existing applications to shaping a blueprint for future ethical AI systems. Our study presents a reproducible method for identifying not only traditional ethical concerns but also new contextual ethical aspects. Incorporating the discovered ethical dimensions into rela-life practices can enhance user trust and satisfaction as well as the long-term acceptance of AI technologies in mental health care. Our future work will extend toward app-specific or longitudinal evaluations to identify how ethical perceptions change over time.

\bibliographystyle{plain}
\bibliography{references}

\end{document}